\documentclass[11pt]{article}
\usepackage{amssymb}
\usepackage{amsmath}
\usepackage{amscd}
\usepackage{latexsym}

\catcode`@=11
\renewcommand{\section}{\@startsection{section}{1}{0pt}{\medskipamount}
{\medskipamount}{\large\bf}}
\numberwithin{equation}{section}
\catcode`@=12
\def\a{\alpha}
\def\b{\beta}

\def\de{\delta}
\def\eps{\epsilon}
\def\ve{\varepsilon}

\def\vp{\varphi}

\newcommand{\Cbb}{\mathbb C}
\newcommand{\R}{\mathbb R}

\newcommand{\Gcal}{{\cal G}}
\newcommand{\Acal}{{\cal A}}
\newcommand{\Ach}{{\widehat{\cal A}}}

\newcommand{\Ical}{{\cal I}}
\newcommand{\Mcal}{{\cal M}}
\newcommand{\Fcal}{{\cal F}}
\newcommand{\Fch}{{\widehat{\cal F}}}
\newcommand{\Ncal}{{\cal N}}

\newcommand{\Pcal}{{\cal P}}

\newcommand{\gfrak}{{\mathfrak g}}
\newcommand{\mfrak}{{\mathfrak m}}
\newcommand{\hfrak}{{\mathfrak h}}

\newcommand{\bari}{{\bar{\imath}}}
\newcommand{\hati}{{\hat{\imath}}}
\newcommand{\barj}{{\bar{\jmath}}}
\newcommand{\hatj}{{\hat{\jmath}}}

\def\tr{\textrm{tr}}
\def\diff{\textrm{d}}
\def\pa{\mbox{$\partial$}}
\def\sfrac#1#2{{\textstyle\frac{#1}{#2}}}
\def\+{\dagger}
\def\={\ =\ }
\def\und{\qquad\textrm{and}\qquad}
\def\and{\quad\textrm{and}\quad}
\def\with{\quad\textrm{with}\quad}
\def\for{\quad\textrm{for}\quad}

\def\Id{\mathrm{Id}}

\begin{document}

\begin{titlepage}
\setcounter{page}{0}

\hspace{2.0cm}

\begin{center}

{\LARGE\bf
Skyrme-Faddeev model from 5d super-Yang-Mills
}

\vspace{12mm}

{\Large
Olaf Lechtenfeld${}^{\+\times}$ \ and \  Alexander D. Popov${}^\+$
}\\[10mm]
\noindent ${}^\+${\em
Institut f\"ur Theoretische Physik,
Leibniz Universit\"at Hannover \\
Appelstra\ss{}e 2, 30167 Hannover, Germany
}\\
{Email: alexander.popov@itp.uni-hannover.de}
\\[5mm]
\noindent ${}^\times${\em
Riemann Center for Geometry and Physics,
Leibniz Universit\"at Hannover \\
Appelstra\ss{}e 2, 30167 Hannover, Germany
}\\
{Email: olaf.lechtenfeld@itp.uni-hannover.de}

\vspace{20mm}

\begin{abstract}
\noindent We consider 5d Yang--Mills--Higgs theory with a compact ADE-type gauge group $G$ and one adjoint scalar field on
$\R^{3,1}\times\R_+$, where $\R_+=[0,\infty )$ is the half-line. The maximally supersymmetric extension of this model, with five
adjoint scalars, appears after a reduction of 6d $\Ncal{=}\,(2,0)$ superconformal field theory on $\R^{3,1}\times\R_+\times S^1$
along the circle~$S^1$. We show that in the low-energy limit, when momenta along $\R^{3,1}$ are much smaller
than along $\R_+$, the 5d Yang--Mills--Higgs theory reduces to a nonlinear sigma model on $\R^{3,1}$ with a coset $G/H$
as its target space. Here $H$ is a closed subgroup of $G$ determined by the Higgs-field asymptotics at infinity.
The 4d sigma model describes an infinite tower of interacting fields, and in the infrared it is dominated by the 
standard two-derivative kinetic term and the four-derivative Skyrme--Faddeev term.

\end{abstract}

\end{center}
\end{titlepage}

\section {Introduction and summary}

\noindent Quantum chromodynamics (QCD) as well as Yang--Mills theory are strongly coupled in the infrared limit,
and hence  the perturbative expansion for them breaks down. In the absence of a quantitative  understanding
of non-perturbative QCD, convenient
alternatives at low energy are provided by effective models among which nonlinear sigma models play an important role.
A first model of this kind was introduced by Skyrme~\cite{1} for describing baryons as point-like solitons (see e.g.~\cite{2}
for a review and references). The standard Skyrme model encodes pion degrees of freedom with an SU(2)-valued function on~$\R^{3,1}$. 
Its action contains the standard two-derivative sigma-model term as well as the four-derivative Skyrme term which stabilizes solitons
against scaling.

A related model was introduced by Faddeev~\cite{3}. This is a sigma model on $\R^{3,1}$ with a coset space $S^2\,{=}\,$SU(2)$/$U(1) as its
target space, and it also contains a four-derivative Skyrme-type term. 
Static Skyrme--Faddeev solitons are maps from $\R^3\cup\{\infty\}=S^3$ to the target space $S^2$ 
and thus characterized by their homotopy class, the Hopf invariant.
The cores of Skyrme--Faddeev solitons, sometimes called Hopfions, are twisted and knotted circles, in contrast to
point-like cores of Skyrmions~\cite{4}-\cite{8}.
It is believed that the Skyrme model and its extension to other mesons provides a low-energy description of baryons,
and that the Skyrme--Faddeev model may describe glueballs~\cite{9} or stable closed vortices in various areas of physics
(see e.g.~\cite{10,11,12} and references therein).
Both the Skyrme and the Skyrme--Faddeev model have been generalized to an arbitrary compact Lie group $G$ and coset $G/H$,
respectively (see e.g.~\cite{13, 14, 15}).

A classical problem of the standard Skyrme model was its difficulty to incorporate other mesons besides pions.
This shortcoming was overcome recently with an extended 4d Skyrme model obtained from 5d Yang--Mills theory 
derived from D-brane configurations in string theory and the holographic approach~\cite{16}
(see e.g.~\cite{17,18,19} for reviews and references). This extended Skyrme model can also be reached from 
6d $\Ncal{=}\,(2,0)$ superconformal field theory compactified on a circle to 5d super-Yang--Mills (SYM) theory 
on $\R^{3,1}\times\Ical$, where $\Ical=[-R, R]$ is a finite-length interval~\cite{20}, upon forgetting the five adjoint scalar fields.

Here, we show that, like the extended Skyrme model, also an extended 4d Skyrme--Faddeev model can emerge in a low-energy limit
of 5d SYM theory with Dirichlet boundary conditions~\cite{21,22}.
In contrast to the extended Skyrme model, for the extended Skyrme--Faddeev model one needs to keep one of the five adjoint scalars
and also to modify the fifth dimension from $\Ical=[-R, R]$ to the half-line $\R_+=[0, \infty)\ni x^4$.
The boundary conditions required for the reduction to $\R^{3,1}$ are encoded in Nahm equations along the fifth dimension~\cite{21,22}.
In our case, the latter become ``baby'' Nahm equations on $\R_+$ for the remaining adjoint scalar~$\phi\in\gfrak =\,$Lie$\,G$.
Solutions to these equations were studied in~\cite{23}.
The scalar is taken to approach an element~$\tau$ of the Cartan subalgebra of $\gfrak$ in the limit $x^4\to\infty$.
The moduli space~$\Mcal_\tau$ of solutions to the baby Nahm equation then becomes the adjoint orbit of~$\tau$. In other words,
$\Mcal_\tau=G/H=\{g\,\tau g^{-1}\,|\,g\in G\}$, where $H$ is the stabilizer of $\tau$ in~$G$.
This coset $G/H$ becomes the target space for our 4d effective sigma model.

We start with 5d SYM theory on $\R^{3,1}\times\R_+$ and show how an extended Skyrme--Faddeev model appears rather naturally
in the low-energy limit.
Our derivation employs the adiabatic approach~\cite{24}-\cite{30} based on Manton's seminal paper~\cite{24}. 
This might give a clue to the construction of a 4d supersymmetric Skyrme--Faddeev model, which seems to have not yet been completed. 
To this end one should keep three of the five adjoint scalars obeying Nahm equations on~$\R_+$. 
To summarize, we demonstrate that not only the Skyrme model but also the Skyrme--Faddeev model as well as their extended versions
emerge from the M5-brane system of M-theory.

\section {Yang--Mills and Higgs fields in five dimensions}

\noindent {\bf Gauge fields and adjoint scalars.} Let $M^d$ be an oriented smooth manifold
of dimension $d$, $G$ a compact ADE-type Lie group with $\gfrak$  its Lie algebra,
$P$ a principal  $G$-bundle over $M^d$, $\Acal$ a connection one-form on $P$
and $\Fcal =\diff \Acal + \Acal\wedge\Acal$ its curvature. We consider also the bundle
of groups Int$P=P\times_G G$ ($G$ acts on itself by internal automorphisms $h\mapsto ghg^{-1}$ with $h, g\in G$)
associated with $P$ and the bundle of Lie algebras Ad$P=P\times_G \gfrak$ (adjoint action of
$G$ on $\gfrak$). These associated bundles inherit their connection $\Acal$ from $P$. Besides $\Acal$
we will also consider $\gfrak$-valued scalar fields $\phi$ on $M^d$, they are sections of the bundle
Ad$P$.

We denote by $\Gcal$ the infinite-dimensional
group of gauge transformations,
\begin{equation}\label{2.1}
\Gcal \ni f:\ \Acal\ \mapsto\ \Acal^{f}= f^{-1} \Acal\, f + f^{-1}\diff f \quad\und\quad
\phi\ \mapsto\ \phi^f=f^{-1}\phi\, f \ ,
\end{equation}
which can be identified with the space of global sections of the bundle Int$P$.
Correspondingly, the infinitesimal action  of $\Gcal$ is defined by global sections $\eps$
of the bundle Ad$P$,
\begin{equation}\label{2.2}
{\rm Lie}\,\Gcal \ni\eps :\  \de_\eps\Acal \=D_{\Acal}\eps\=\diff\eps + [\Acal,\eps ]
\quad\und\quad \de_\eps \phi\=[\phi , \eps]\ .
\end{equation}
The moduli space of pairs $(\Acal ,\phi )$ is defined as the quotient of the space of all
such pairs by the action (\ref{2.2}) of the gauge group $\Gcal$.

\medskip

\noindent {\bf Space $\R^{3,1}\times\R_+$.} We consider $d{=}5$ and Yang--Mills--Higgs theory on
$M^5=\R^{3,1}\times\R_+$ with coordinates $(x^{\mu})=(x^a, x^4)$ for $a=0,1,2,3$,
where $x^a\in \R^{3,1}$ and $x^4\in\R_+=[0, \infty )$.
We introduce a family of flat metrics,
\begin{equation}\label{2.3}
 \diff s^2_\ve \= g_{\mu\nu}^\ve\,\diff x^\mu \diff x^\nu \=
\eta_{ab}\,\diff x^a \diff x^b + \ve^2(\diff x^4)^2\ ,
\end{equation}
where $(\eta_{ab})={\rm diag} (-1,1,1,1)$ and $\ve >0$ is a dimensionless parameter
regulating the transition to the low-energy limit. Namely, for $\ve =1$ one has the standard Yang--Mills--Higgs
theory on $\R^{3,1}\times\R_+$.
For small $\ve$, momenta along $\R_+$ are much larger than momenta along $\R^{3,1}$, and Yang--Mills--Higgs
theory on $\R^{3,1}\times\R_+$ reduces to a non-linear sigma model on $\R^{3,1}$ that will be described below.

Note that the definition of infrared limit hiddenly introduces an arbitrary scale~$L$ into the model.
In five dimensions this scale is provided by $e^2$, where $e$ is the (dimensionful) 5d gauge coupling.
For physical application it is to be matched, e.g.~to the nuclear scale.
Here, the infrared region is defined by $\ve\ll 1$ for convenience. 
For a dimensionful variant, one may absorb the length dimension of~$x^4$ into~$\ve$ and 
take the infrared domain as $\ve\ll L$.

\medskip

\noindent {\bf Action functional.} For a $\gfrak$-valued gauge potential (connection) $\Acal$ and its
gauge field  (curvature) $\Fcal$ on the principal bundle
$P$ over $\R^{3,1}\times\R_+$ we have the obvious splitting
\begin{equation}\label{2.4}
\Acal \= \Acal_{a}\,\diff x^a+\Acal_{4}\,\diff x^4\und
\Fcal \=\sfrac12\Fcal_{ab}\,\diff x^a \wedge \diff x^b + \Fcal_{a4}\,\diff x^a \wedge \diff x^4\ .
\end{equation}
The fields $\Acal$, $\Fcal$ and $\phi$ are taken in the adjoint representation of $\gfrak =\,$Lie$G$. 
For the adjoint generators $I_i$ of $G$ we use the standard
normalization $\tr (I_iI_j) = -2\de_{ij}$ with $i,j=1,\ldots,{\rm dim}G$.

For the metric tensor (\ref{2.3}) we have $(g^{\mu\nu}_\ve)=(\eta^{ab}, \ve^{-2})$ and
$\det(g_{\mu\nu}^\ve)=-\ve^{2}$. We denote by $\Fcal^{\mu\nu}_\ve$
the contravariant components raised from $\Fcal_{\mu\nu}$ by  $g^{\mu\nu}_\ve$ and by
$\Fcal^{\mu\nu}$ those obtained by using $g^{\mu\nu}\equiv g^{\mu\nu}_{\ve =1}$.
We have $\Fcal^{ab}_\ve=\Fcal^{ab}$ and $\Fcal^{a4}_\ve=\ve^{-2}\Fcal^{a4}$. We also rescale
the Higgs field $\phi\mapsto \ve^{-1}\phi$.
The Yang--Mills--Higgs (YMH) action functional on $\R^{3,1}\times \R_+$ with the metric  (\ref{2.3})
then takes the form
\begin{equation}\label{2.5}
\begin{aligned}
S&\=-\frac{1}{8e^2}\int_{\R^{3,1}\times \R_+} \!\!\!\!\!\!\diff^5x\
\sqrt{|\det g^\ve|}\ \tr\left (\Fcal_{\mu\nu}\Fcal^{\mu\nu}_\ve + \sfrac2{\ve^2} D_{\mu}\phi D^{\mu}\phi\right ) \\
&\= -\frac{1}{8e^2}\int_{\R^{3,1}\times \R_+} \!\!\!\!\!\!\diff^5x\
\tr\bigl( \ve\,\Fcal_{ab}\Fcal^{ab}+\sfrac2\ve\Fcal_{a4}\Fcal^{a4}+\sfrac2\ve D_{a}\phi D^{a}\phi+
\sfrac{2}{\ve^3} D_{4}\phi D_{4}\phi\bigr)\ .
\end{aligned}
\end{equation}
There is no potential for $\phi$ in (\ref{2.5}), as in
the standard action for monopoles. Instead, nontrivial geometry for $\phi$ will appear from asymptotic
conditions at infinity. The action (\ref{2.5}) follows from the bosonic action of maximally
supersymmetric Yang--Mills theory in five dimensions (see e.g.~\cite{22}) after putting to zero 
four of the five adjoint scalar fields.

\bigskip

\section{Boundary conditions and moduli space of vacua}

\noindent {\bf Conditions at $x^4=0$ and at $x^4\to\infty$.} 
For convenience let us introduce a dimensionless fifth coordinate~$z$ and dimensionless field components,
\begin{equation}
\label{rescale}
z \= x^4/L \und \Acal_z \= L\,\Acal_{x^4} \qquad\textrm{as well as}\qquad \vp \=L\,\phi\ .
\end{equation}
The boundary of $M^5=\R^{3,1}\times\R_+$ consists of Minkowski
space $\R^{3,1}_0=\pa M^5$ at $z=0$.
Infinity $z\to\infty$ is parametrized by Minkowski space $\R^{3,1}_\infty$ at $z=\infty$.
For the $\gfrak$-valued fields $(\Acal , \vp )$ on $\R^{3,1}\times\R_+$ 
we have to impose boundary (at $z=0$)
and asymptotic (for $z\to\infty$) conditions. 
We make the following choice~\cite{30a},
\begin{equation}\label{3.1}
\Acal_a(x^a, z{=}0)=0\und 
\bigl\{ \pa_z\vp(x^a, z)+[\Acal_z(x^a, z),\vp(x^a, z)]\bigr\}\big|_{z=0}=0\ ,
\end{equation}
\begin{equation}\label{3.2}
\Acal_\mu (x^a, z{\to}\infty)=0\und \vp(x^a, z{\to}\infty)=\tau(x^a)\in\mathfrak t \subset \mathfrak g\ ,
\qquad\qquad\qquad\ \ {}
\end{equation}
where $\mathfrak t$ is a Cartan subalgebra of the Lie algebra $\gfrak$. 
Such a class of conditions is parametrized by the a Minkowski-space $\gfrak$-valued function $\tau$
and has been imposed e.g.~in studies of the Nahm equations on $\R_+$ (see e.g.~\cite{31,32,23}).

\medskip

\noindent {\bf Gauge group.} We employ the notation $\Ical :=\R_+$ and consider the group
$\Gcal = C^\infty (\R^{3,1}\times\Ical, G$) as well as its restriction $\Gcal_\Ical$ to $\Ical$
obtained by fixing $x^a\in\R^{3,1}$ to some arbitrary value, i.e. $\Gcal_\Ical\cong C^\infty (\Ical ,G)$.
The true group of gauge transformations has to preserve the chosen boundary and asymptotic conditions
(\ref{3.1}) and (\ref{3.2}) (see e.g.~\cite{33}). 
This is not the case for~$\Gcal$ but for its subgroup
\begin{equation}\label{3.5a}
\Gcal^0\=\bigl\{h\in\Gcal :\quad h(x^a,z{=}0)=h(x^a,z{\to}\infty )=\Id\bigr\}\ .
\end{equation}
In the following, we shall need two larger subgroups, which preserve (\ref{3.1}) but not the
asymptotics~(\ref{3.2}), namely
\begin{equation}\label{3.3}
\Gcal^1 \= \bigl\{ h\in \Gcal :\quad h(x^a, z{=}0)=\Id \quad\textrm{but}\quad h(x^a, z{\to}\infty)\in G\bigr\} \und
\end{equation}
\begin{equation}\label{3.4}
\Gcal^\tau \= \bigl\{ h\in \Gcal :\quad h(x^a,z{=}0)=\Id \and h(x^a, z{\to}\infty)\in H\bigr\}\ ,\qquad\qquad\quad\!{}
\end{equation}
where $H$ is the stabilizer of $\tau$ in $\mathfrak t$ under the adjoint action. 
Clearly, $\Gcal^0\subset\Gcal^\tau\subset\Gcal^1\subset\Gcal$, and the transformations from $\Gcal^\tau$
respect the asymptotics~(\ref{3.2}) only for $\Acal_z$ and $\vp$.

For the Lie algebras $\gfrak$ and $\hfrak$ of the Lie groups $G$ and $H$, respectively, we have $\gfrak=\hfrak\oplus\mfrak$
and choose $\mfrak$ to be orthogonal to $\hfrak$ with respect to the Cartan--Killing form.
We assume that the adjoint orbit $G/H$ is reductive, which means that $[\hfrak , \mfrak ]\subset \mfrak$. When $H$
is the maximal torus $T$ in $G$, the coset space $G/T$ is the orbit of maximal dimension.

We denote by $\Gcal^1_\Ical$ and $\Gcal^\tau_\Ical$ the restrictions of the groups  $\Gcal^1$ and $\Gcal^\tau$
to the half-line $\Ical=\R_+$ by fixing $x^a\in\R^{3,1}$. It follows from the definitions of $\Gcal_\Ical$, $\Gcal^1_\Ical$
and $\Gcal^\tau_\Ical$ that
\begin{equation}\label{3.5}
\Gcal_\Ical/\Gcal^1_\Ical\ \cong\ G \und \Gcal^1_\Ical/\Gcal^\tau_\Ical\ \cong\ G/H
\end{equation}
since the elements of these groups differ only either at $z=0$ or at $z=\infty$.
Correspondingly, the definitions of $\Gcal$, $\Gcal^1$ and $\Gcal^\tau$ imply that
\begin{equation}\label{3.6}
\Gcal/\Gcal^1\ \cong\ C^\infty (\R^{3,1}, G) \und \Gcal^1/\Gcal^\tau\ \cong\ C^\infty (\R^{3,1}, G/H)\ .
\end{equation}

\medskip

\noindent {\bf Yang--Mills--Higgs model on $\R_+$.} Our consideration of the low-energy limit $\ve\to 0$ of the YMH
model (\ref{2.5}) is based on the adiabatic approach which for YMH theories was introduced in the
seminal paper~\cite{24} by Manton (for brief reviews see e.g.~\cite{30, 34} and references therein).
In the adiabatic approach one should firstly restrict the YMH theory (\ref{2.5}) to $\Ical$ and classify solutions on
$\Ical$ not depending on $x^a\in\R^{3,1}$ and secondly declare that their moduli, which parametrize such solutions,
depend on $x^a\in\R^{3,1}$ and derive the effective action for these moduli functions.

From the action (\ref{2.5}) it follows that for $\ve\to 0$ and $\Acal_a=0$ the equations of motion read
\begin{equation}\label{3.7}
\pa_a\Acal_z=0=\pa_a\vp\und \pa_z\vp+[\Acal_z,\vp ]=0\ .
\end{equation}
The conditions (\ref{3.1}) and (\ref{3.2}) become
\begin{equation}\label{3.8}
\Acal_z(z{=}\infty)=0  \und
\vp(\infty) =\tau\in {\mathfrak t}\subset\gfrak\ ,
\end{equation}
while the boundary condition~(\ref{3.1}) at $z{=}0$ is satisfied due to (\ref{3.7}).
For regular elements~$\tau$ (when $H$ is the maximal torus $T$ in $G$), solutions to (\ref{3.7}) and (\ref{3.8})
were described in~\cite{23}. We adapt the construction to non-regular~$\tau$.

Equation~(\ref{3.7}) is solved by
\begin{equation}\label{3.9}
\Acal_z=h^{-1}\pa_zh\und \vp= h^{-1}\vp(0)\,h \qquad\textrm{where}\quad h(z)\in \Gcal_\Ical\ .
\end{equation}
However, $h(z)$ and $h^{-1}(0)h(z)$ define the same solution, so we may impose $h(0)=\Id$ or, equivalently,
take $h(z)\in \Gcal_\Ical^1$. Then from (\ref{3.8}) and (\ref{3.9}) we obtain that
\begin{equation}\label{3.10}
\vp(\infty) = h^{-1}(\infty)\,\vp(0)\, h(\infty)=\tau \qquad \Rightarrow\qquad
\vp(0) = h(\infty)\,\tau\,h^{-1}(\infty)\ .
\end{equation}
As $H$ is the stabilizer of~$\tau$ under the adjoint $G$-action, 
\begin{equation}\label{3.11}
h^{}_0\,\tau\, h_0^{-1} =\tau\for h^{}_0\in H\ ,
\end{equation}
we may locally factorize 
\begin{equation}\label{3.12}
h(\infty)=m\,h_0 \qquad\Rightarrow\qquad \vp(0) = m\,\tau\, m^{-1}\in G/H \quad
\for h_0\in H \and m\in G/H\ ,
\end{equation}
so that $\Mcal_\Ical = G/H$ is the moduli space of solutions to (\ref{3.7}).

\medskip

\noindent {\bf Moduli space of vacua.} One arrives at the same vacuum moduli space $\Mcal_\Ical = G/H$
for YMH theory on $\R_+$ by noting the one-to-one correspondence between $\Acal_z$
on $\R_+$ and $h(z)\in\Gcal_\Ical^1$ given by the first formula in (\ref{3.9}) and its ``inverse"
\begin{equation}\label{3.13}
h(z)\=\Pcal\exp\bigl (\smallint^z_0 \Acal_y\diff y\bigr )\ ,
\end{equation}
where $\Pcal$ denotes path ordering. The gauge subgroup $\Gcal^\tau_\Ical$ acts on $\Acal_z$ and $\vp (z)$ 
(and hence on the solution space $\Gcal^1_\Ical\ni h(z)$) by
\begin{equation}\label{3.14}
\Gcal^\tau_\Ical \ni f:\quad \Acal_z \mapsto\ \Acal^{f}_z= f^{-1} \Acal_z  f + f^{-1}\pa_z f \ ,
\quad \vp\mapsto \vp^f=f^{-1}\vp\,f \quad\Rightarrow\quad h\mapsto h^f=hf\ .
\end{equation}
Hence, the moduli space of solutions (\ref{3.9}) is
$\Mcal_\Ical=\Gcal^1_\Ical/\Gcal^\tau_\Ical\cong G/H$, and one can define the principal
$\Gcal^\tau_\Ical$-bundle
\begin{equation}\label{3.15}
q:\quad\Gcal^1_\Ical\stackrel{\Gcal^\tau_\Ical}{\longrightarrow} G/H \quad\with h(z)\mapsto m
\end{equation}
for $m\in G/H$ defined in (\ref{3.12}).

\bigskip

\section{Infinitesimal change of solutions $(\Acal_z,\vp)$}

\noindent {\bf Linearized equations.} Suppose we have a solution $(\Acal_z, \vp )$ to 
(\ref{3.7}), which belongs to the moduli space $\Mcal_\Ical =G/H$ from (\ref{3.15}). Then
$(\de\Acal_z, \de\vp )$ will be a tangent vector to $G/H$ at the point $(\Acal_z, \vp )$ if
\begin{equation}\label{4.1}
D_z\de\vp + [\de\Acal_z, \vp]=0
\end{equation}
and
\begin{equation}\label{4.2}
D_z\de\Acal_z + [\vp , \de\vp]=0\ ,
\end{equation}
where $D_z=\pa_z+[\Acal_z , \cdot \ ]$. Equation (\ref{4.1}) means that $(\de\Acal_z, \de\vp )$ belong
to the tangent space $T^{}_{(\Acal_z, \vp )}\Gcal^1_\Ical$ of the solution space $\Gcal^1_\Ical$,
and (\ref{4.2}) says that $(\de\Acal_z, \de\vp )$ is orthogonal to the gauge modes (cf.~\cite{26} for
a similar discussion regarding the moduli space of monopoles in~$\R^3$). Below we will explain this in more
detail.

\medskip

\noindent {\bf Geometry of $G/H$}. We consider the adjoint orbit (\ref{3.12}). Let us choose a basis $\{I_i\}$
for the Lie algebra $\gfrak$ in such a way that $\{I^{}_{\bari}\}$ for $\bari =1,\ldots,\dim G/H$ form a basis
for $\mfrak$ and $\{I^{}_{\hati}\}$ for $\hati =\dim G/H +1,\ldots,\dim G$ provide a basis for $\hfrak$.
For the total Lie algebra we have $\gfrak=\hfrak\oplus\mfrak$ and $\tr (I^{}_{\bari}I^{}_{\hati})=0$.

The space $G/H$ consists of left cosets $gH$, and the natural projection $g\mapsto gH$ is denoted by
\begin{equation}\label{4.3}
\pi :\quad G \stackrel{H}{\longrightarrow} G/H\ .
\end{equation}
On $G/H$ there exists an orthonormal frame of left-invariant one-forms $\{e_{}^{\bari}\}$ which locally provides
the $G$-invariant metric
\begin{equation}\label{4.4}
\diff s^2_{G/H}\=\de_{\bari\barj}\,e^\bari e^\barj \= \de_{\bari\barj}\,e^\bari_\a e^\barj_\b\, \diff X^\a \diff X^\b
\ =:\ g_{\a\b}\,\diff X^\a\diff X^\b \quad\for \a,\b=1,\ldots,\dim G/H\ ,
\end{equation}
where $X^\a$ are local coordinates on $G/H$. The principal $H$-bundle (\ref{4.3})
supports a unique $G$-invariant connection, the so called {\it canonical connection\/}~\cite{35,36,37},
\begin{equation}\label{4.5}
\Acal_{G/H}=e^\hati I_\hati = e^\hati_\bari\,I_\hati\,e^\bari=e^\hati_\a I_\hati\,\diff X^\a\ .
\end{equation}
The one-forms $e^i=(e^\bari , e^\hati )$ obey the Maurer--Cartan equations,
\begin{equation}\label{4.7}
\diff e^\bari =-f^{\bari}_{\hatj\bar{k}}\, e^\hatj\wedge e^{\bar k} - \sfrac12\,f^\bari_{\barj\bar k}\, e^\barj\wedge e^{\bar k} \und
\diff e^\hati =-\sfrac12\,f^{\hati}_{\hatj\hat{k}}\, e^\hatj\wedge e^{\hat k} - \sfrac12\,f^\hati_{\barj\bar k}\, e^\barj\wedge e^{\bar k}\ .
\end{equation}
The curvature of the canonical connection (\ref{4.5}) follows as
\begin{equation}\label{4.8}
\Fcal_{G/H}\=  - \sfrac12\,f^\hati_{\barj\bar k}\, I_{\hati}\, e^\barj\wedge e^{\bar k}
\= - \sfrac12\,f^\hati_{\barj\bar k}\, I_{\hati}\, e^\barj_\a e^{\bar k}_\b\, \diff X^\a\wedge\diff X^\b\ .
\end{equation}

\medskip

\noindent {\bf Variation of $(\Acal_z, \vp )$}. Recall that the solution space $\Gcal^1_\Ical$ to (\ref{3.7}) is a group,
and the moduli space $\Mcal_\Ical=\Gcal^1_\Ical/\Gcal^\tau_\Ical$ is labelled locally by coset coordinates $X=\{X^\a\}$.
Let us pick a coset representative $m(X)\in\Gcal^1_\Ical$, which is a section of the bundle~(\ref{3.15}) 
over a point~$X\in G/H$. Multiplication from the left by a group element $h\in \Gcal^1_\Ical$ will generally 
carry $m(X)$ into a section~$m(X')$ over another point~$X'$, so that
\begin{equation}\label{4.9}
h\,m(X) = m(X')\,f \quad\with f\in\Gcal^\tau_\Ical\ .
\end{equation}
This yields formulae for the infinitesimal changes of $\Acal_z$ and $\vp$, 
which live in Lie$\,\Gcal^1_\Ical=\mfrak\,\oplus\,$Lie$\,\Gcal^\tau_\Ical$,
\begin{equation}\label{4.10}
\pa_\a\Acal_{z}\=\de_{\a}\Acal_{z} + \de_{\eps_\a} A_z \= \de_{\a}\Acal_{z} + D_z\eps_\a \und 
\pa_\a\vp\=\de_\a\vp + \de_{\eps_\a}\vp \= \de_\a\vp + [\vp ,\eps_\a]\ ,
\end{equation}
where $\pa_\a=\pa/\pa X^\a$. The pair $(\de_\a\Acal_z, \de_\a\vp )$ belongs to the tangent space
$T_{(\Acal_z, \vp )} \Mcal_\Ical\cong \mfrak$,
and $\eps_\a$ are $\gfrak$-valued gauge parameters generating the infinitesimal gauge transformation
$(\de_{\eps_\a} A_z,\de_{\eps_\a}\vp)$ which represents the gauge part of the variation and
sits in Lie$\,\Gcal^\tau_\Ical$.
The orthogonality of  $(\de_\a\Acal_z, \de_\a\vp )$ and
$(\de_{\eps_\a} A_z , \de_{\eps_\a}\vp )$ is achieved by imposing the condition
(\ref{4.2}) for any $\a = 1,\ldots,\dim G/H$.

\bigskip

\section{Skyrme--Faddeev model in the infrared limit of 5d YMH}

\noindent {\bf Coset space sigma model.} 
We return to the YMH model (\ref{2.5}) on $\R^{3,1}\times \R_+$ and non-vacuum fields $(\Acal_a,\Acal_z,\vp)$. 
The adiabatic approach considers the collective coordinates $X=\{X^\a\}$ as dynamical fields, $X^\a = X^\a (x)$,
where $x=\{x^a\}$.
Their low-energy effective action is derived by expanding
\begin{equation}\label{5.1}
\Acal_{\mu} \= \Acal_{\mu}\bigl(X(x),z\bigr) + \ldots \und \vp \= \vp \bigl(X(x),z\bigr) + \ldots
\end{equation}
and keeping only the first terms in the YMH action (\ref{2.5})~\cite{24,25,26,28,20}.
Thereby one obtains an effective field theory which will be a non-linear sigma model describing maps
$X:\R^{3,1}\to G/H$.

With the map~$X$ we pull back the adiabatic fields
\begin{equation}
\Acal_z\=\Acal_z\bigl(X(x), z\bigr)\ =:\ \Acal_z(x,z) \und 
\vp \=\vp\bigl(X(x), z\bigr)\ =:\ \vp(x,z)
\end{equation}
by a slight abuse of notation from $G/H$ to $\R^{3,1}$.
Thus, we have to include a dependence on $x^a$ in the formulae of Sections 3 and~4.
In particular, multiplying (\ref{4.10}) by $\pa_a X^\a$, we obtain
\begin{equation}\label{5.2}
\pa_a \Acal_{z} \= (\pa_a X^\a)\de_\a\Acal_z +D_z\eps_a \und 
\pa_a\vp \= (\pa_a X^\a)\de_{\a}\vp+[\vp , \eps_a]\ ,
\end{equation}
where $\eps_a= (\pa_a X^\a)\,\eps_{\a}$  is the pull-back of $\eps_\a$ to $\R^{3,1}$.
From (\ref{5.2}) it follows that
\begin{equation}\label{5.3}
\Fcal_{az}\=\pa_a\Acal_z - D_z\Acal_a\= (\pa_a X^\a)\de_\a\Acal_z - D_z(\Acal_a{-}\eps_a)\ ,
\end{equation}
\begin{equation}\label{5.4}
D_a\vp \=\pa_a\vp +[\Acal_a, \vp]\= (\pa_a X^\a)\de_\a\vp - [\vp , \Acal_a{-}\eps_a]\ .
\end{equation}
In the moduli-space approximation, $\Fcal_{a4}$ and $D_a\vp$ are tangent to $\Mcal_{\Ical}$
(see e.g.~\cite{24,25,26}). This can be achieved by putting
\begin{equation}\label{5.5}
\Acal_a\=\eps_a \bigl(X(x), z\bigr)\ .
\end{equation}
Then, substituting (\ref{5.3}) and (\ref{5.4}) into the action (\ref{2.5}) and remembering~(\ref{rescale}), 
we arrive at
\begin{equation}\label{5.6}
S_{\textrm{kin}}\=-\frac{1}{4 e^2 \ve}\int_{\R^{3,1}\times\R_+} \!\!\!\!\!\diff^5x\ \eta^{ab}\
\tr\left(\Fcal_{a4}\Fcal_{b4}+D_a\phi D_b\phi\right)
\=\frac{1}{2e^2\ve L}\int_{\R^{3,1}} \!\!\diff^4x\ \eta^{ab}\,g_{\a\b}\,\pa_a X^\a \pa_bX^\b \ ,
\end{equation}
where
\begin{equation}\label{5.7}
g_{\a\b}\=-\sfrac12\int^{\infty}_{0}\!\diff z\ \tr\left(\de_\a \Acal_z\de_\b \Acal_z+\de_\a \vp\de_\b\vp \right)
\= \de_{\bari\barj}\, e^\bari_\a e^\barj_\b
\end{equation}
are the components of the metric (\ref{4.4}) on $G/H$ pulled-back to $\R^{3,1}$. Thus, this part of the action
(\ref{2.5}) reduces to the standard non-linear sigma model on $\R^{3,1}$ with the coset $G/H$ as target space.

\medskip

\noindent {\bf Skyrme--Faddeev term}.
The last term in the action (\ref{2.5}) vanishes since $D_4\phi =0$ for any $x^a\in\R^{3,1}$ due to
second equation in (\ref{3.7}). It remains to evaluate the first term in the action~(\ref{2.5}).
For this we notice that
$\Acal_a=\eps_a=(\pa_aX^\a)\eps_\a$ depend on $X^\a(x)$ and $z$ and that
\begin{equation}\label{5.8}
\eps_a (z{=}0)=0 \und \eps_a (z{=}\infty)\in \hfrak\ .
\end{equation}
The  asymptotics (\ref{5.8}) at $z\to\infty$ does not agree with the asymptotic conditions (\ref{3.2})
for the components $\Acal_a$.
The reason is that, when we turn from YMH theory on $\R_+$ to YMH theory on $\R^{3,1}\times \R_+$, 
the group of gauge transformations are reduced from $\Gcal^\tau$ to $\Gcal^0$.
To preserve (\ref{3.2}) we switch from $\Acal_a$ to $\Ach_a$ via
\begin{equation}\label{5.9}
\Acal_a \= \eps_a \= f^{-1}\,\Ach_a\,f + f^{-1}\,\partial_a f \qquad\textrm{with some}\quad f\in\Gcal^\tau\ .
\end{equation}
The conditions (\ref{5.8}) for $\Acal_a$ translate to
\begin{equation}\label{5.10}
\Ach_a(z{=}0) = 0 \und \Ach_a(z{=}\infty) = 0
\end{equation}
since $f(z{=}0) =0$ and $f(z{=}\infty)\in H$ and $\Acal_a(z{=}\infty)=f^{-1}\,\partial_a f\in \hfrak$.

Recall that 
\begin{equation}
\Ach_a\=(\pa_aX^\a)\,\hat\eps_\a \qquad\textrm{where}\qquad
\eps_\a \= f^{-1}\,\hat\eps_\a\,f + f^{-1}\,\partial_\a f\ .
\end{equation}
This $\hat\eps_\a\,\diff X^\a$ is a one-form on the base $G/H$ of the fibration with value in
the Lie algebra Lie$\,\Gcal^0$. One can always decompose  $\hat\eps_\a$ as
\begin{equation}\label{5.11}
\hat\eps_\a \= \zeta (z)\,e^\hati_\a\, I_\hati +\eps^0_\a\ ,
\end{equation}
where $A_\a =  e^\hati_\a\, I_\hati $ are the components of the unique $G$-equivariant connection (\ref{4.5}) 
in the bundle  (\ref{4.3}),
and $\zeta (z)$ is a real-valued function on $\R_+$ such that $\zeta (0)=0=\zeta (\infty)$.
One can view (\ref{5.11}) as a definition of $\eps_\a^0$. Then for $\Ach_a$ we have
\begin{equation}\label{5.12}
\Ach_a\=\zeta (z) (\pa_aX^\a ) e^\hati_\a\, I_\hati + (\pa_aX^\a )\eps_\a^0\=\zeta (z)A_a +\eps_a^0\ .
\end{equation}
We remark that $A_a$ is a composite field since the canonical connection  (\ref{4.5}) has a fixed dependence on
the coordinates $X^\a$ on $G/H$, which is known explicitly if one chooses 
$G/H=\;$SU$(n{+}1)/$U$(n)$, SU$(n{+}1)/$U$(1)^n$ or similar.
On the other hand, $\eps_a^0$ is not composite.

The curvature of  $\Ach$ computes to
\begin{equation}\label{5.13}
f\Fcal f^{-1} \=\Fch\=\diff\Ach + \Ach\wedge \Ach  \= F + \Sigma\  ,
\end{equation}
where
\begin{equation}\label{5.14}
F\=\zeta\,\diff A + \zeta^2 A\wedge A \=\sfrac12\,F_{ab}\,\diff x^a\wedge\diff x^b \und \qquad\quad{}
\end{equation}
\begin{equation}\label{5.15}
\Sigma\=\diff \eps^0 +\eps^0\wedge\eps^0 + \zeta( A\wedge \eps^0+\eps^0\wedge A) \= \sfrac12\,\Sigma_{ab}\,\diff x^a\wedge\diff x^b\  .
\end{equation}
We will see in a moment that the term $F$ in  (\ref{5.13}) yields a Skyrme--Faddeev type term for a generic coset space $G/H$. On the other hand,
the curvature $\Sigma$ in (\ref{5.13}) describes $\gfrak$-valued one-forms with non-vanishing mass terms from the coupling to the composite
field~$A$ in~(\ref{5.15}). A consideration of these fields is beyond the scope of our paper, which contents itself with identifying the
Skyrme--Faddeev model as part of the low-energy limit of 5d SYM on $\R^{3,1}\times\R_+$.
The discarded term $\Sigma$ will yield corrections analogous to the tower of meson fields in the extended Skyrme model (see e.g.~\cite{16,19}).

For the components $F_{ab}$ from (\ref{5.14}) and (\ref{4.5})-(\ref{4.7}) we obtain
\begin{equation}\label{5.16}
F_{ab}\=\bigl(\zeta(\zeta{-}1)f^{\hati}_{\hatj\hat{k}}\,e^\hatj_\a e^{\hat k}_\b-\zeta\,f^\hati_{\barj\bar k}\,e^\barj_\a e^{\bar k}_\b\bigr)\,
I_\hati\ \pa_aX^\a  \pa_bX^\b \ .
\end{equation}
Substituting  (\ref{5.13}) into the action  and discarding all $\Sigma_{ab}$ terms, the first term in  (\ref{2.5}) produces
\begin{equation}\label{5.17}
\begin{aligned}
S_{\textrm{SF}} &\=
-\frac{\ve}{8 e^2}\int_{\R^{3,1}\times \R_+} \!\!\!\!\diff^5x\ \tr\Fcal_{ab}\Fcal^{ab} \=
\frac{\ve L}{4 e^2} \int_{\R^{3,1}} \!\diff^4x\ 
\eta^{ac}\eta^{bd}\ \pa_aX^\a \pa_bX^\b\pa_cX^\gamma\pa_dX^\de\ \times \\[4pt] &\qquad\qquad\times\,\left\{ 
a_1 f^\hati_{\hat l\hat k}f^\hatj_{\hat m\hat n}\,e^{\hat l}_\a  e^{\hat k}_\b e^{\hat m}_\gamma e^{\hat n}_\de +
a_2 f^\hati_{\hat l\hat k}f^\hatj_{\bar m\bar n}\,e^{\hat l}_\a  e^{\hat k}_\b e^{\bar m}_\gamma e^{\bar n}_\de +
a_3 f^\hati_{\bar l\bar k}f^\hatj_{\bar m\bar n}\,e^{\bar l}_\a  e^{\bar k}_\b e^{\bar m}_\gamma e^{\bar n}_\de \right\} \de_{\hati\hatj}\ ,
\end{aligned}
\end{equation}
with numerical coefficients
\begin{equation}\label{5.18}
a_1=\int_0^\infty\diff z\ \zeta^2(\zeta{-}1)^2\ ,\qquad a_2=\int_0^\infty\diff z\ \zeta^2(\zeta{-}1)\und a_3=\int_0^\infty\diff z\ \zeta^2
\end{equation}
The integrals (\ref{5.18}) are finite for a suitably chosen function $\zeta(z)$ such as $\zeta(z)=\exp (-z)(1-\exp (-z))$.
The expression~(\ref{5.17}) for the Skyrme--Fadeev-type term holds true for generic cosets $G/H$.
It considerably simplifies when $H=T$ is the Cartan torus in $G$, because then $f^\hati_{\hatj\hat k}=0$,
and one has only the $a_3$ term in~(\ref{5.17}). For $G/T=\;$SU(2)$/$U(1), this term coincides with the standard
Skyrme--Faddeev term of the $\Cbb P^1$ sigma model. 

To summarize, in the infrared limit the Yang--Mills--Higgs action (\ref{2.5}) on $\R^{3,1}\times \R_+$
is reduced to the effective action of the Skyrme--Faddeev model
\begin{equation}\label{5.19}
S_{\textrm{eff}} \= S_{\textrm{kin}} + S_{\textrm{SF}} \ ,
\end{equation}
where $S_{\textrm{kin}}$ and $S_{\textrm{SF}}$ are given by (\ref{5.6}), (\ref{5.7}) and (\ref{5.17}).

\bigskip

\noindent {\bf Acknowledgements}

\noindent
This work was partially supported by the Deutsche Forschungsgemeinschaft grant LE 838/13.
It is based upon work from COST Action MP1405 QSPACE, supported by COST (European Cooperation
in Science and Technology).

\newpage

\noindent {\bf Note added after review}

\noindent
By similar methods, the authors recently obtained the {\sl standard\/} 4d Faddeev and Skyrme models in an infrared limit of 4d Yang--Mills--Higgs theory.
Breaking the gauge group $G$ to a subgroup $H$ results in a Higgs vacuum manifold $G/H$, which coincides with the Faddeev sigma-model target.
The coset may be chosen to be a group manifold, e.g.~$G/H\simeq\textrm{U}(N)$, in which case the standard $\textrm{U}(N)$ Skyrme model emerges~\cite{38}.


\bigskip

\begin{thebibliography}{99}
\addtolength{\itemsep}{-6pt}

\bibitem{1}
T.H.R.~Skyrme,
 ``A unified field theory of mesons and baryons,''
  Nucl.\ Phys.\  {\bf 31} (1962) 556.

\bibitem{2}
 I.~Zahed and G.E.~Brown,
  ``The Skyrme model,''
  Phys.\ Rept.\  {\bf 142} (1986) 1.

\bibitem{3}
L.~Faddeev, ``Quantization of solitons,'' 
IAS preprint, Print-75-QS70, Princeton, 1975.

\bibitem{4}
L.D.~Faddeev and A.J.~Niemi,
  ``Knots and particles,''
  Nature {\bf 387} (1997) 58
  [hep-th/9610193].

\bibitem{5}
L.D.~Faddeev and A.J.~Niemi,
  ``Toroidal configurations as stable solitons,''
  hep-th/9705176.

\bibitem{6}
R.A.~Battye and P.M.~Sutcliffe,\\
  ``Knots as stable soliton solutions in a three-dimensional classical field theory,''\\
  Phys.\ Rev.\ Lett.\  {\bf 81} (1998) 4798
  [hep-th/9808129].

\bibitem{7}
J.~Hietarinta and P.~Salo,
``Faddeev--Hopf knots: Dynamics of linked unknots,''\\
  Phys.\ Lett.\ B {\bf 451} (1999) 60
  [hep-th/9811053].

\bibitem{8}
P.~Sutcliffe,
``Knots in the Skyrme--Faddeev model,''\\
  Proc.\ Roy.\ Soc.\ Lond.\ A {\bf 463} (2007) 3001
  [arXiv:0705.1468 [hep-th]].

\bibitem{9}
L.D.~Faddeev and A.J.~Niemi,
  ``Partially dual variables in SU(2) Yang--Mills theory,''\\
  Phys.\ Rev.\ Lett.\  {\bf 82} (1999) 1624
  [hep-th/9807069].

\bibitem{10}
E.~Babaev, L.D.~Faddeev and A.J.~Niemi,\\
  ``Hidden symmetry and knot solitons in a charged two-condensate Bose system,''\\
  Phys.\ Rev.\ B {\bf 65} (2002) 100512
  [cond-mat/0106152 [cond-mat.supr-con]].

\bibitem{11}
E.~Radu and M.~S.~Volkov,
 ``Existence of stationary, non-radiating ring solitons in field theory: knots and vortons,''
  Phys.\ Rept.\  {\bf 468} (2008) 101
  [arXiv:0804.1357 [hep-th]].

\bibitem{12}
L.D.~Faddeev,
  ``Knots as possible excitations of the quantum Yang--Mills fields,''\\
 in: Proc. Conf. in Honor of C.N.~Yang's 85th Birthday,
{\it Statistical Physics, High Energy, Condensed Matter and Mathematical Physics},
eds. M.-L. Ge, C.H. Oh and K.K. Phua (World Scientific, Singapore, 2008) p.18
  [arXiv:0805.1624 [hep-th]].

\bibitem{13}
D.~Auckly and L.~Kapitanski,
  ``Holonomy and Skyrme's model,''\\
  Commun.\ Math.\ Phys.\  {\bf 240} (2003) 97
  [math-ph/0211010].

\bibitem{14}
L.D.~Faddeev and A.J.~Niemi,
``Partial duality in SU(N) Yang--Mills theory,''\\
  Phys.\ Lett.\ B {\bf 449} (1999) 214
  [hep-th/9812090].

\bibitem{15}
 L.A.~Ferreira and P.~Klimas,
  ``Exact vortex solutions in a $CP^N$ Skyrme--Faddeev type model,''
  JHEP {\bf 10} (2010) 008
  [arXiv:1007.1667 [hep-th]].

\bibitem{16}
T.~Sakai and S.~Sugimoto,
 ``Low energy hadron physics in holographic QCD,''\\
  Prog.\ Theor.\ Phys.\  {\bf 113} (2005) 843
  [hep-th/0412141].

\bibitem{17}
 J.~Erdmenger, N.~Evans, I.~Kirsch and E.~Threlfall,
 ``Mesons in gauge/gravity duals -- a review,''
  Eur.\ Phys.\ J.\ A {\bf 35} (2008) 81
  [arXiv:0711.4467 [hep-th]].

\bibitem{18}
 V.~Kaplunovsky, D.~Melnikov and J.~Sonnenschein,
  ``Baryonic popcorn,''\\
  JHEP {\bf 11} (2012) 047
  [arXiv:1201.1331 [hep-th]].

\bibitem{19}
P.~Sutcliffe,
  ``Holographic skyrmions,''
  Mod.\ Phys.\ Lett.\ B {\bf 29} (2015) 1540051.

\bibitem{20}
T.A.~Ivanova, O.~Lechtenfeld and A.D.~Popov,
 ``Skyrme model from 6d $\cal N$=(2,0) theory,''
  Phys.\ Lett.\ B {\bf 783} (2018) 222
  [arXiv:1805.07241 [hep-th]].

\bibitem{21}
D.~Gaiotto, G.W.~Moore and Y.~Tachikawa,
 ``On 6d $\mathcal N=$(2,0) theory compactified on a Riemann surface with finite area,''
  PTEP {\bf 2013} (2013) 013B03
  [arXiv:1110.2657 [hep-th]].

\bibitem{22}
B.~Assel, S.~Sch\"afer-Nameki and J.M.~Wong,
  ``M5-branes on $S^{2}\times M_{4}$: Nahm's equations and 4d topological sigma-models,''
  JHEP {\bf 09} (2016) 120
  [arXiv:1604.03606 [hep-th]].

\bibitem{23}
R.~Bielawski, ``Lie groups, Nahm's equations and hyper-K\"ahler manifolds,''\\
 in: {\it Algebraic Groups}, Universit\"atsverlag G\"ottingen, G\"ottingen, 2007 [math/0509515].

\bibitem{24}
 N.S.~Manton,
  ``A remark on the scattering of BPS monopoles,''
  Phys.\ Lett.\ B {\bf 110} (1982) 54.

\bibitem{25}
  J.A.~Harvey and A.~Strominger,
  ``String theory and the Donaldson polynomial,''\\
  Commun.\ Math.\ Phys.\ {\bf 151} (1993) 221
  [hep-th/9108020].

\bibitem{26}
J.P.~Gauntlett,
  ``Low-energy dynamics of N=2 supersymmetric monopoles,''\\
  Nucl.\ Phys.\ B {\bf 411} (1994) 443
  [hep-th/9305068].

\bibitem{27}
  S.K.~Donaldson and R.P.~Thomas, ``Gauge theory in higher dimensions,''\\
  in: {\it The Geometric Universe}, Oxford University Press, Oxford, 1998.

\bibitem{28}
  N.S.~Manton and P.~Sutcliffe, {\it Topological solitons},\\
  Cambridge University Press, Cambridge, 2004.

\bibitem{29}
  O.~Lechtenfeld and A.D.~Popov,
  ``Yang--Mills moduli space in the adiabatic limit,''\\
  J.\ Phys.\ A {\bf 48} (2015)  425401
  [arXiv:1505.05448 [hep-th]].

\bibitem{30}
  T.A.~Ivanova, O.~Lechtenfeld and A.D.~Popov,
  ``Non-Abelian sigma models from Yang--Mills theory compactified on a circle,''
Phys.\ Lett.\ B {\bf 781} (2018)  322
  [arXiv:1803.07322 [hep-th]].
  
\bibitem{30a}
  D.~Gaiotto and E.~Witten,
  ``Supersymmetric boundary bonditions in ${\cal N}=4$ super Yang--Mills theory,''
  J.\ Statist.\ Phys.\  {\bf 135} (2009) 789
  [arXiv:0804.2902 [hep-th]].

\bibitem{31}
P.B.~Kronheimer, ``A hyper-K\"ahlerian structure on coadjoint orbits of a semisimple complex group,''
J.\ London Math.\ Soc.\ {\bf 42} (1990) 193.

\bibitem{32}
A.G.~Kovalev, ``Nahm's equations and complex adjoint orbits'',\\
Quart.\ J.\ Math.\ Oxford {\bf 47} (1996) 41.

\bibitem{33}
  S.K.~Donaldson,
  ``Boundary value problems for Yang--Mills fields,''\\
  J.\ Geom.\ Phys.\ {\bf 8} (1992) 89.

\bibitem{34}
A.~Deser, O.~Lechtenfeld and A.D.~Popov,
  ``Sigma-model limit of Yang--Mills instantons in higher dimensions,''
  Nucl.\ Phys.\ B {\bf 894} (2015) 361
  [arXiv:1412.4258 [hep-th]].

\bibitem{35}
S.~Kobayashi and K.~Nomizu, {\it Foundations of differential geometry}, vol.1,\\ 
Interscience Publishers, New York, 1963.

\bibitem{36}
J.P.~Harnad, J.~Tafel and S.~Shnider,
 ``Canonical connections on Riemannian symmetric spaces and solutions to the Einstein--Yang--Mills equations,''
 J.\ Math.\ Phys.\ {\bf 21} (1980) 2236.

\bibitem{37}
D.~Harland, T.A.~Ivanova, O.~Lechtenfeld and A.D.~Popov,\\
  ``Yang--Mills flows on nearly K\"ahler manifolds and $G_2$-instantons,''\\
  Commun.\ Math.\ Phys.\  {\bf 300} (2010) 185
  [arXiv:0909.2730 [hep-th]].

\bibitem{38}
  O.~Lechtenfeld and A.D.~Popov,
  ``Skyrme and Faddeev models in the low-energy limit of 4d Yang--Mills--Higgs theories,''
  arXiv:1808.08972 [hep-th].

\end{thebibliography}
\end{document}